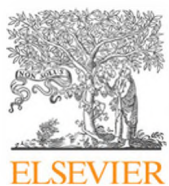

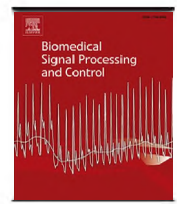

# Enhanced average for event-related potential analysis using dynamic time warping

Mario Molina [a], Lorenzo J. Tardón [a,*], Ana M. Barbancho [a], Irene De-Torres [b], Isabel Barbancho [a,*]

[a] *Universidad de Málaga, ATIC Research Group, Campus de Teatinos sn, Málaga, 29071, Málaga, Spain*
[b] *Hospital Regional Universitario de Málaga, Av. de Carlos Haya, 84, Málaga, 29010, Málaga, Spain*

**ARTICLE INFO**

**ABSTRACT**

Electroencephalography (EEG) provides a way to understand, and evaluate neurotransmission. In this context, time-locked EEG activity or event-related potentials (ERPs) are often used to capture neural activity related to specific mental processes. Normally, they are considered on the basis of averages across a number of trials. However, there exist notable variability in latency jitter, jitter, and amplitude, across trials, and, also, across users; this causes the average ERP waveform to blur, and, furthermore, diminish the number of underlying waves.

For these reasons, a strategy is proposed for obtaining ERP waveforms based on dynamic time warping (DTW) to adapt, and adjust individual trials to the averaged ERP, previously calculated, to build an enhanced average by making use of these warped signals. At the sight of the experiments carried out on the behaviour of the proposed scheme using publicly available datasets, this strategy reduces the attenuation in amplitude of ERP components thanks to the reduction of the influence of variability of latency and jitter, and, thus, improves the averaged ERP waveforms.

## 1. Introduction

Event-related potentials, or evoked-response potentials (ERPs), are widely used in neuroscience to relate specific mental processes to some stimulus: visual, auditory, etc. These are voltage fluctuations associated to synchronous events that can be extracted from electroencephalography (EEG) signals [1]. An example of direct application of ERPs is the brain–computer interface (BCI) context, which aims at providing a direct communication path between the brain and the external environment [2–4].

Within this context, analysis procedures are based on a certain stimulus that is repeated a number of times; then the recorded EEG signals are conveniently averaged to diminish the perturbation caused by sources of noise and other unrelated brain signals to produce a representative waveform. Note that although the averaging scheme can be considered from the perspective of the presence of multiple closely space cognitive processes [5], this approach is commonly considered for short stimuli or isolated events.

This common strategy for obtaining ERPs, based on synchronous averaging individual EEG trials corresponding to the same isolated stimulus, is able to reduce the effects of random EEG signal variations, ongoing brain activity or 'noise' across the event-related observations on each trial [1] to achieve a representation of the waveform linked to the stimulus [6–8]. This EEG response may present peaks, and troughs that are denoted by names like N100 or P300 [9], or, within the ERP analysis context of early sensory processing, P50 [10].

Regarding the simple averaging strategy, in [11], the authors review several aspects related to ERPs, and their utilization in the diagnoses of cognitive aspects of certain clinical disorders. In [1], the authors also consider the averaging problem, and review, and propose alternatives to conventional averaging based on weighting, and trimming the conventional measure.

A different approach is followed by Samar et al. in [12], where the utilization of wavelets for the analysis of ERP data is considered. Other strategies can be examined; in [9], Canonical Correlation Analysis (CCA) is employed to eliminate brain response variance unrelated to stimuli, specifically oriented to the continuous stimuli scenario; in [13], ERP temporal variations are considered from the feature extraction point of view to intervene in classification and detection tasks.

This kind of approaches, and considerations highlights the fact that the assumption of stationarity of ERPs is at the core of the averaging procedure [14]. But latency jitter [15], or delay jitter [16] and non-linear timing distortions of signals (jitter) [17] affect the averaging process, and can cause the reduction of amplitude of important ERP components [18].

In this context, Woody, in [19], proposes an alternative to the traditional stimulus-locked ERP average to cope with trial-by-trial latency differences on the basis of cross-correlations between single-trial waveforms, and a template averaged waveform: single-trial waveforms are time shifted according to the time lag given by the maximum of the cross-correlation. But, no temporal variability across trials is considered.

In [17], the authors consider a warping scheme by employing an underlying model of temporal displacement and alignment based on least squares. Gibbons and Stahl, [20], consider the correction of response time for ERP averaging too: the authors assume that significant variations in the timing of ERP components, especially in late components, must be expected; so they propose the usage of a polynomial expression for response time correction which is based on the previous identification of a mean response time and linear interpolation. The iterative use of averaging DTW without taking into account, noise, jitter or the occurrence of new frequencies is presented in [21]. In [22], the author proposed a modification of a cost matrix that can diminish the unfavourable results in the suppression of noise when averaging of non-linearly aligned signal cycles.

In [14], the authors explicitly state that ERPs can suffer significant across-trial variability in terms of latency and latency jitter that may lead to important distortion of the averaged ERP and, thereby, to the misestimation of their amplitude and latency. They review some methods to deal with these problems, including the usage of Fourier transform, independent component analysis (ICA) or time frequency decomposition based on continuous wavelet transform (CWT), though these are not specifically oriented to deal with jitter and latency jitter issues.

Fukami et al. in [23], consider the problem of variability between trials of ERPs related to a certain stimulus taking into account the possible influence of background EEG and noise; they focus on the P300 component, and, regarding our specific context, a fixed scalar latency is considered by the authors to cope with.

In [16], the generation model of ERPs is observed, and a procedure to evaluate two different models on the basis of properties of the components found is considered. With that, the authors deal with jitter and latency jitter, though they do not aim to correct them or process the signals to build an average, but to identify an underlying model.

Other works have considered the utilization of DTW, explicitly aiming at providing or facilitating the calculation of a certain measurement; consequently, they are not focused on obtaining a template or representative ERP signal. In [24], clustering of EEG data artificially generated is presented, but there is no modification of the DTW approach to make the results usable to deal with averages. In a different scenario, DTW is used in [25] to align sequences for the analysis of similarity between a signal template and individual signals. However, the aim is it not to build the template, but to guide a resampling scheme for the later utilization of the correlation coefficient to measure similarity. On the other hand, in [26], the authors use DTW to measure the dissimilarity distances of EEG signals with a KNN classifier.

In [27], the authors compare an adaptative DTW barycentre averaging method [28], to avoid the problem of length of the resulting average sequence, but its application is in an accelerometer-based hand gesture recognition system. DTW has been also used in EEG-based emotion recognition as similarity measure of brain rhythm sequencing [29]. Also, single-channel selection for emotion recognition in EEG signals using brain rhythm sequencing used DTW for sequence classification [30].

In [31], the authors compare DTW against Fisher–Rao Registration to study the phase variability; they highlight the problems associated to DTW regarding the modification of signals, and subsequent corruption of features in them, which limits its utilization, and, specifically, making the original DTW scheme unusable for averaging.

Within this context, a new method is proposed in this manuscript to obtain the ERP waveform using modified dynamic time warping (DTW) for short or isolated stimuli. A modification of the DTW method is proposed to stretch and synchronize individual trials to a reference signal, the averaged ERP, thus leading to a new averaged ERP waveform. As in the conventional strategy, the stimulus must be repeated several times giving rise to a number of trials for the calculation of their average. This averaged signal will be used as reference to align individual signals using the proposed modified DTW scheme. These lead, by further averaging, to a new averaged ERP. Also, a low-pass filter is applied to the aligned signals to reduce the effect of spurious samples caused by the DTW alignment process. Following this approach, the experiments show that the impact of jitter is reduced as well as the influence of noise, thus improving the ERP components in terms of their amplitude, specially.

Note that our approach differs from the one in [17], where data normalization is required to avoid the degradation of results due to possible isolated large square errors, and their approach to warping is completely different from the point of view of the definition of a cost matrix, and whole warping procedure. Also, the work presented in [20] might be the most similar approach to our proposal; however, our scheme does not impose a polynomial restriction over the temporal variable, being more flexible, and, also, our scheme does not require previous identification of the response time of selected events, or components, or specific amplitude interpolation schemes.

The performance of the proposed scheme will be evaluated in terms of root-mean-square RMS deviation and maximum absolute difference (MAD) between the individual trials and the different averaged ERP signals, as well as by means of the observation of characteristics of the P200 component, and a basic example of a classification scheme.

This manuscript is organized as follows: in Section 2, event-related potentials (ERP), and dynamic time warping (DTW) are briefly exposed. In Section 3, the databases employed to build examples and evaluations, the proposed enhanced ERP averaging scheme, and the selected quantitative evaluation measures are described. In Section 4, results, and comparisons between different approaches to obtain average ERP waveforms are presented, and discussed; also a simple signal classification example is drawn to illustrate the usability of signal models obtained with the proposed scheme in such context. Finally, conclusions are summarized in Section 5.

## 2. Event-related potentials and dynamic time warping

This section briefly revises the concept of event-related potential (ERP), as well as the original dynamic time warping algorithm to align two signals, on which our modified scheme is based.

### 2.1. Event-related potentials (ERPs)

Event-related potentials (ERPs), or evoked-response potentials are time-locked neural signals related to specific events or stimuli.

ERPs are EEG signals of low amplitude, (between 4 to 10 μV), that are considered to reflect joint activity of postsynaptic potentials when numerous neurons fire synchronously [32]. Note that, though some actions can be considered [33], noise, and other perturbations are unavoidable. Within this context, the utilization of averages across responses corresponding to the same activities, and state help to reduce the influence of noise with respect to the underlying common signal. However, other issues like delay or jitter still affect the recovery of such signal [14–16]. Precisely, our proposal is oriented to fight against noise, delay, and jitter.

Within the framework of ERP signals, different waveforms are relevant regarding their analysis [32]. For example, the positive wave (positive deflection: peak) with a latency period around 300 ms (in [32], the interval to find such wave is considered to be between 250 and 400 ms for adult subjects between 20 and 70 years), called P300 (or simply P3) [34], is considered by some authors to be an index of the decision-making process [35]. Regarding these waveforms, several elements can be identified that can modify the observed P300 (or other P and N waves) [15,32,36]:

- Jitter: random variation of delay across time.
- Latency jitter: the latency of the wave is the time measured from the stimulus onset to the peak (or negative deflection: valley) which appears within certain time ranges. Some authors argue that this latency reflects the stimulus evaluation time, thus, the stimuli evaluation time is assumed to vary depending on the difficulty of the response discrimination [37], which may lead to significant across-trial variability of response latency or latency jitter [14].
- Overlapping components: for example, regarding P300, Feedback Related Negativity (FRN) or Error-Related Negativity (ERN) occur at similar timing, resulting in reduced positivity.

Another positive component is the visual P200 or P2, measured in the centre-frontal and parieto-occipital area of the cerebral cortex, between 150 and 275 milliseconds. This is also related to visual responses, language information, memory, and repetitive effects [38]. A negative peak, called N100, is usually found between 80 and 120 milliseconds after a stimulus, and distributed over the fronto-central region of the head. It usually appears in auditory stimuli, although if also arises in relation to other stimuli, such as visual ones [7,39].

It should be clear, at this point, that detecting ERPs is an important task. Since the ERP is usually a response to a stimulus with a relatively stable amplitude and shape, the most commonly used analysis methods consist of repeating a stimulus many trials, and averaging the time-locked EEG signals. Each analysis epoch is normally formed by a pre-stimulus, stimulus, and post-stimulus intervals [40].

Note that, in order to use the ERP averaging method, signal pre-processing, filtering or other techniques to reduce or remove artefacts (like eye motion, heartbeat or movement) are required. Also, three conditions must be satisfied [40]:

- The signal is time-locked.
- The noise is not correlated with the signal.
- Latency is relatively stable, and jitter is low.

Difficulties arise when averaging ERPs due to variability between trials in terms of latency, jitter, and amplitude. Precisely, the averaged ERP waveform actually comes from a mixture of waveforms with amplitude variations, and different degrees of latency fluctuations and jitter [18].

When the latency of a component varies throughout trials (latency or jitter variations), the average ERP becomes blurry, and, so, its waveform is affected. Also, the higher ERP amplitude and latency variations, the greater ERP waveform variability [41].

## 2.2. Dynamic time warping (DTW)

Traditionally, DTW has been used in speech and sound processing to achieve accurate synchronization between audio excerpts [42]. It is a dynamic programming technique that accommodates differences in time between two signals [43].

The algorithm is based on allowing a number of predefined steps in the space of time, or samples, to find a path that maximizes the local coincidence between aligned signals [43,44]. So, this optimal path is found as the one with the shortest distance, according to a certain distance measure, between each pair of points complying with a set of rules or conditions, specifically [45,46]:

- Each element of the first signal must match one or more elements of the other signal, and vice versa.
- The first element of the first signal must match, at least, the first element of the other.

- The last element of the first signal must coincide, at least, with the last element of the other.
- The mapping of elements of the first signal to elements of the other one must be increasing, and conversely.

The total cost can be measured as the sum of absolute differences (or the selected distance measure) between the values of each pair of coincident elements. It serves as a measure of agreement between the signals [47].

## 3. Materials and processing methods

This section describes, first, the datasets of EEG signals used for the description, and illustration of the behaviour of the proposed processing scheme; then, the EEG pre-processing stage, and the proposed DTW-based averaging scheme are exposed; finally, the measures selected for quantitative evaluation and comparison of schemes are drawn.

### 3.1. Dataset 1: Simanova et al.

This EEG dataset was obtained from [48]. These data correspond to a study that investigates, on the basis of the analysis of ERPs, the semantic processing of stimuli presented as images, visualized text or words shown in an auditory way. These data have been stored, and organized following BIDS (Brain Imaging Data Structure): a standard brain imaging data structure for organizing and describing neuroimaging tests [49].

*Experiment and subjects.* In the study, 24 native Dutch people (10 men and 14 women) participated. Their age range was between 18 and 28 years. All of them were right-handed; also none of them had suffered any psychological or neurological illness. Those experiments were approved by a local ethics committee, and all subjects gave their written consent, as described in [48].

*Stimuli.* The stimuli presented to build the EEG dataset can be divided into two semantic categories, animals and tools. There are 4 items in each category:

- Animals: cow, bear, lion and ape.
- Tools: axe, scissors, comb and pen.

For this database, a third category with clothing and vegetables is also considered, but it was not used in the experiments and tests performed in our case.

Each item is presented to the subjects in three different stimulus modalities:

- Auditory: spoken in Dutch.
- Visual: drawings in black on a white background.
- Textual: words written in Dutch in black on a white background

Text and image stimuli were presented for 300 ms, followed by a black screen lasting between 1000 and 1200 ms, then the next stimulus is shown, and the process is repeated until its conclusion. For auditory stimuli, the interval was also 1000 and 1200 ms, and a cross was displayed on the screen meanwhile [48].

For the specific tests used in this manuscript, all the categories (tools and animals), and modalities (spoken, written and picture) were employed.

*Task performed.* The research in [48] was aimed at applying the Bayesian approach to identify concept-related neuronal activity on the basis of ERP analysis. To this end, subjects participating in the experiment were instructed to classify each stimulus presented into two irrelevant categories (clothing or vegetables) in order to make the subjects categorize those stimuli without openly conducting the assessment between actual relevant classes. The subjects define their response by pressing a button. Additional specific details can be found in [48].

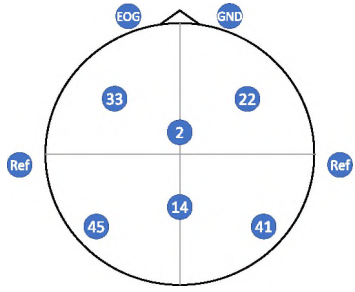

**Fig. 1.** Distribution of the EEG electrodes (channels) used. *Ref* identifies the reference, *EOG* stands for the electrode that measures electrooculogram, and *GND* corresponds to ground.

*EEG recording.* EEG data were acquired with a BrainProducts BrainAmp EEG amplifier with 64 channels (filtered between 0.2 and 200 Hz, and sampled at 500 Hz) using BrainVision Recorder Professional software. The electrode cap used holds 60 electrodes placed on the scalp. EEG data were recorded against the reference at the right mastoid; an additional electrode measured the voltage on the left mastoid, and the data were offline converted to a linked-mastoids reference. Vertical and horizontal bipolar electrooculogram (EOG) channels were computed after acquisition, using electrodes placed horizontally and vertically around the eyes [48].

Recall that this paper is focused on the behaviour of the proposed DTW-based averaging scheme, similarly to other works like [1,5], or [50]; therefore, out of the 64 recorded channels [48], in this study, channels 2, 22, 33, 14, 41 and 45 have been selected in order to cover different regions of the brain; other pieces of data in the dataset were analysed, though results are not presented here, since they do not provide any further comprehension or information regarding the behaviour of the scheme proposed. Fig. 1, shows the distribution of the EEG electrodes (channels) selected.

### 3.2. Dataset 2: Birbaumer et al.

This EEG dataset has been made available by University of Tuebingen [51,52]. It contains slow cortical potentials for healthy subjects, and artificially respirated amyotrophic lateral sclerosis (ALS) patients. This EEG dataset has been chosen to illustrate the usability of the proposed DTW-based scheme in a simple classification context in relation to BCI, and neurological disease.

*Experiment and subjects.* Four pieces of the dataset are available [51, 52]; among them, the subsets with labels for classification are employed. This means that subsets 'traindata_1a' and 'traindata_1b' are finally chosen; these correspond to a healthy subject, and an artificially respirated amyotrophic lateral sclerosis (ALS) patient [53].

*Task performed.* The subjects were asked to interact with a computer by moving a cursor up and down on a screen while cortical potentials were measured. The subjects received visual feedback of their slow cortical potentials Cz-Mastoids from second 2 of each trial, which last 3.5, and 5.5 s more for the healthy and ALS subjects, respectively. No rest was scheduled between trials.

*EEG recording.* The recording system employed is a PsyLab EEG8 [54], and the sampling rate was 256 Hz [51]. With reference to the 10/20 system [55], the following EEG channels were recorded: A1-Cz (where A1 = left mastoid), A2-Cz, 2 cm frontal of C3, 2 cm parietal of C3, 2 cm frontal of C4, and 2 cm parietal of C4.

'Traindata_1a' contains 268 trials, with 135 trials belonging to class *Up* and 133 to class *Down*; 'traindata_1b' contains 200 trials, with 100 trials corresponding to each class.

The channel located 2 cm frontal of C3, which shows the best behaviour to illustrate the utilization of the proposed DTW-based averaging scheme for classification among the available channels, is selected, though a different choice could have been done.

### 3.3. EEG pre-processing

Pre-processing of the datasets is carried out following the procedure presented in [48], and [56] using MATLAB R2022a. First, a 30 Hz low-pass FIR filter was applied. Then, further trials were removed to avoid artefacts by visual rejection of data segments based on the variance of epochs by making use of *ft_rejectvisual()*, following [56]. Trials with ocular artefacts or voltage variations above 150 μV, were removed. These trials were not employed for the calculation of ERPs.

*Dataset 1: Simanova et al..* The recorded data, that contains 64 EEG channels, were divided into one-second trials, starting 200 ms before the stimulus (pre-stimulus), and ending 800 ms after it (post-stimulus).

After pre-processing, which includes artefact removal, the total number of trials in each category and modality was just under 400 [48].

*Dataset 2: Birbaumer et al..* Excerpts of 1.2 ms build the trials that are analysed. After pre-processing, a total of 239 and 170 trials remain for the healthy subject and ALS patient cases, respectively.

### 3.4. Enhanced average using DTW

In order to build an improved average, a warped or aligned version using DTW of each trial, for a certain stimulus, is obtained. This signal version represents the one that best fits the common ERP average. However, DTW creates non-linear deformations of the signals with respect to time; this issue needs to be solved to make the scheme usable for averaging in our context. The specific details on the modified DTW algorithm devised for our context are described next.

Let $s_t[n]$ stand for each of the EEG signals corresponding to a channel or component linked to a certain trial, $t$, with $t = 1, \ldots, T$, where $n$ indexes the sampling instants of the trial excerpt $t$: $n = 1 \ldots N$. Recall that these signal excerpts can be defined starting before the stimulus. Also, let $r[n]$ stand for the signal obtained by conventional averaging of selected trials, for a certain subject or across subjects.

The Euclidean distance can be used to build the similarity matrix in a sample-by-sample basis using a multi-feature characterization of each sample; in our case, we employ this distance measure using solely the signal amplitude at each sample for characterization. This means that each element $c_t(i, j)$ in the local cost matrix $C_t$ for a certain trial $t$ is given by:

$$c_t(i, j) = |r[i] - s_t(j)|, \ i, j = 1 \ldots N \tag{1}$$

After the construction of the similarity matrix, the alignment point $p = (p_1, p_2, \ldots, p_M)$ between the two signals must be found, with $p$ representing the sequence of binomials with temporal indexes of the signals: $p_k = (p_i, p_j)$ that describe the alignment path or warping function [46] under specific path conditions:

- The starting and ending points of the warping path must relate the first and last points of each of the aligned sequences: $p_1 = (1, 1)$ and $p_M = (N, N)$.
- The path $p$ is defined by the steps $d_{k-1} = p_k - p_{k-1}$ that advance one unit in the index of either $r[n]$, or $s_t[n]$, or both:

$$d_{k-1} = \begin{cases} (1, 1) \\ (1, 0) \\ (0, 1) \end{cases} \tag{2}$$

After the optimal path is found by following this conventional warping process with the specificities described regarding the similarity matrix, a new version of each trial can be obtained by using this path. However, recall that this process can lead to arbitrarily lengthened signals which would be of no use to calculate the average across trials. In order to avoid this, all the consecutive steps, $d_k$, that do not advance in the reference index for the reconstruction of the warped signal are identified, i.e. $(0, 1)$ steps, to guide the removal of the corresponding path element: $p_{k+1}$. This results in a restricted path $p^*$ that can be used to build a reconstructed signal $s_t^w[n]$ that lasts no longer than the reference.

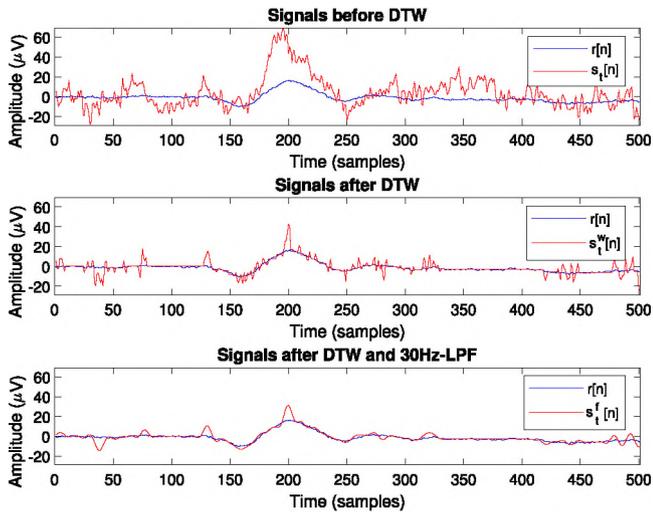

**Fig. 2.** Illustration of the result of the proposed DTW-based alignment process between a signal and its reference. Data used correspond to the first trial of channel 2 of the spoken modality, and the reference signal to the average of channel 2 across the first half of trials from the spoken modality [48].

Note that the reconstructed signal might be shorter than the reference $r[n]$, in that case, the last sample of such reconstructed signal is repeated until the duration of the reference is reached to give the final warped trial signal $s_t^w[n]$. A warped signal $s_t^w[n]$ for each individual trial $t$ is obtained using the reference signal as described. These warped signals can be employed to calculate the DTW-based average $r^w[n]$.

An illustration of the signals obtained after the process described is shown in Fig. 2. It can be observed that the warped signals (middle and bottom figures) get adapted to the reference better than the original one (top figure). In addition, a peak around 200 ms is depicted, with notably greater amplitude in the signal to be aligned ($s_t[n]$) than in the reference signal ($r[n]$). This peak coincides with P200, described before, and could correspond to the processing of language information, since these trials belong to spoken modality data [48].

Observe that the presence of spurious high frequency components is also noticeable; their presence is by no means evenly distributed along $s_t^w[n]$ due to the warping scheme described. This issue can be diminished by applying a filter to constrain the warped signal to the frequency range of the original one. Thus, a minimum-order low-pass finite impulse response (FIR) filter is designed using the Kaiser window method [57] with cut-off frequency equal to the maximum frequency of $s_t[n]$, and stop-band attenuation 60 dB. This step leads to the filtered warped signal $s_t^f[n]$ (bottom image in Fig. 2). The process is described in Fig. 3, where DTW* represents the proposed modified DTW scheme which makes use of $r[n]$ as reference to modify each $s_t[n]$, and LPF represents the low-pass filter with cut-off frequency ($f_M$) defined by the signal under consideration.

These filtered warped signals can then be employed to calculate the filtered DTW-based average $r^f[n]$:

$$r^f[n] = \frac{1}{T} \sum_{t=1}^{T} s_t^f[n] \tag{3}$$

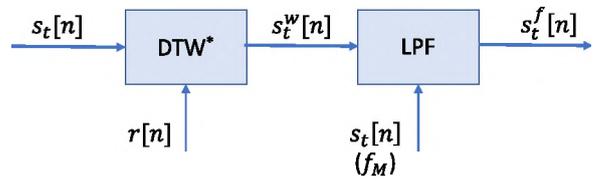

**Fig. 3.** Illustration of the processing scheme to obtain each filtered warped signal $s_t^f[n]$.

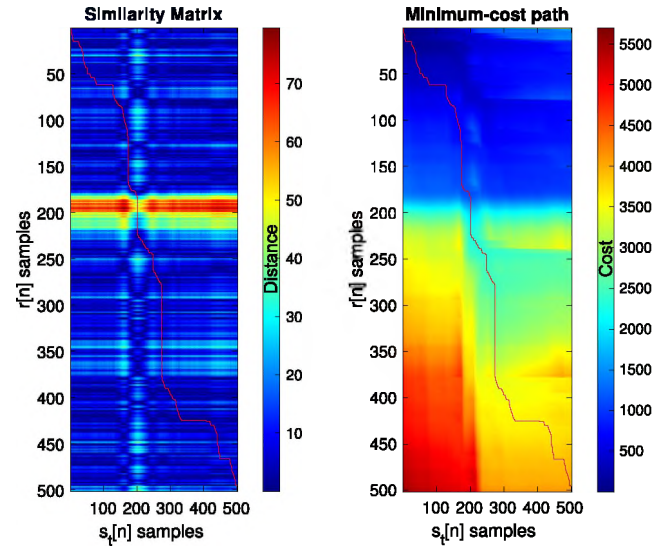

**Fig. 4.** Similarity matrix and minimum-cost path for the modified DTW function with Euclidean distance. Data used belong to the spoken modality of EEG signals provided in [48]. The signal to be aligned corresponds to the first trial of channel 2 of subject 02, and the reference signal to the average of channel 2 across all the spoken modality trials (see Fig. 2).

Fig. 4 displays the similarity matrix, and minimum-cost path found to define the signal $s_t^w[n]$ in Fig. 2. Observe the apparent similarity between the reference and trial signal found at around 200 ms, which is highlighted by the similarity matrix; samples far from that instant (200 ms) in $s_t[n]$ exhibit large dissimilarity with $r[n]$ at around 200 ms clearly due to the presence of P200 in both signals, with samples whose amplitude is notably larger than others in the excerpt.

Fig. 5 illustrates the different path obtained by our modified scheme with respect to the original DTW algorithm on two available speech signals of the same locution with different timing [58]. Spectrogram is employed for signal characterization; cosine distance is used to measure similarity. A solid-red line and a dashed-black line drawn on the similarity matrix, and minimum-cost path matrix represent the optimal path found by the conventional and our modified scheme, respectively.

Observe the difference in optimal path found by both schemes due to the different stages in both algorithms. Specifically, our modified DTW scheme does not allow unconstrained lengthening of the signal to match the reference, avoiding (removing) individual steps in the path that provoke such situation.

### 3.5. Evaluation measures

For the quantitative evaluation of the performance of the considered schemes, in this context, root-mean-square (RMS) value is a commonly used measure [7,17,23]; even when other measures are proposed to gather information from ERPs [50], RMS is considered a main reference. Consequently, RMS is employed as main feature for evaluation of the proposed scheme.

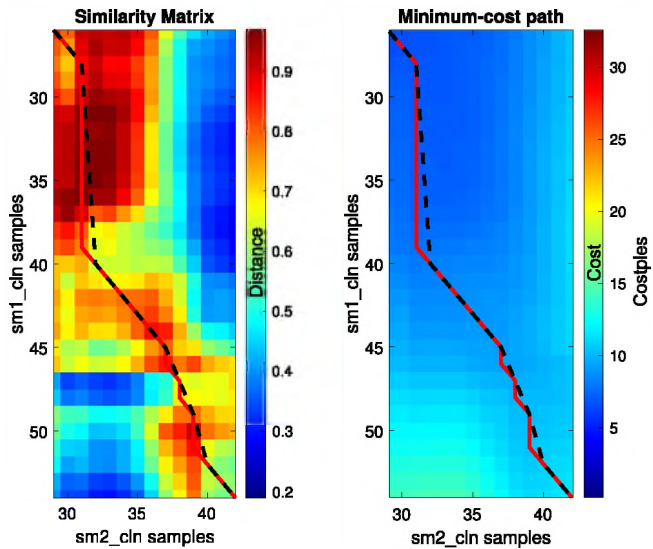

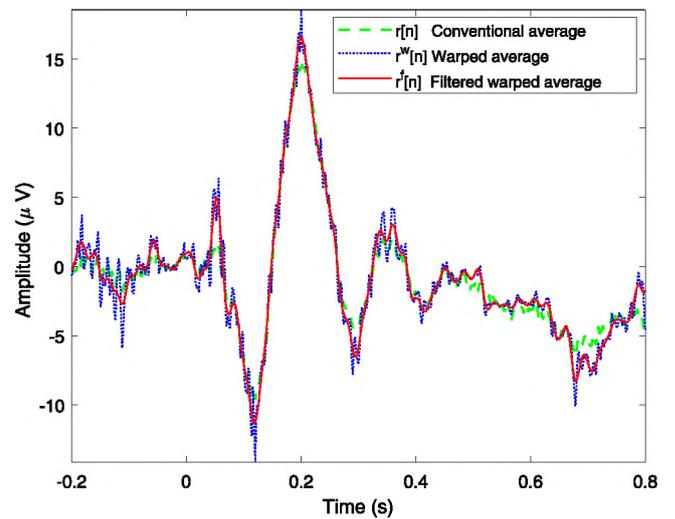

**Fig. 5.** Illustration of the difference between the optimal path matching two speech signals of the same locution but with different timing using the conventional and modified DTW schemes. Specifically, they are the samples 'sm1_cln.wav' and 'sm2_cln.wav', from TIMIT [58], a standard dataset used for the evaluation of speech recognition systems. Optimal path found using our modified DTW scheme is shown in dashed-black line, and the result of the conventional DTW scheme [43] is drawn in solid-red line. Cosine distance is employed to measure similarity between signal representations based on spectrogram.

Root-mean-square (RMS) value between each average: conventional, DTW-based and filtered DTW-based average, and a trial $t$, $s_t[n]$, is calculated as follows:

$$e_t^* = \sqrt{\frac{1}{N} \sum_{n=1}^{N} \left( s_t[n] - r^*[n] \right)^2} \tag{4}$$

where the symbol $*$ is used to refer to conventional signals, when the superscript becomes null, DTW-based signals, when $* = {}^w$, and filtered DTW-based signals, when the superscript is $^f$.

Another evaluation criterion is considered: the maximum absolute difference (it will be denoted MAD) [59,60]. This parameter is a measure of the disparity, absolute difference, between each of the tree considered average signals: $r[n]$, $r^w[n]$ and $r^f[n]$ and a certain trial, $s_t[n]$. It is given by the following expression:

$$MAD_t^*[n] = max_n(abs(s_t[n] - r^*[n])), \tag{5}$$

Also, features of an ERP component will be used for a more complete assessment [23,32].

We will observe the behaviour of P200 delay, peak and amplitude, measured as suggested in [61] in the considered averaging schemes. Intervals to search for peaks and troughs corresponding to N100 and P200 components are defined as described in [32]. The behaviour of these parameters will be analysed per scheme in terms of their mean, standard deviation, coefficient of variation, $V = \frac{STD}{Mean}$ [62], median, 25th quartile, 75th quartile, and maximum and minimum values.

## 4. Results and discussion

In this section, the behaviour and performance of the proposed DTW-based averaging scheme are shown. Additionally, an illustration of the usability of signal models built by the proposed scheme in a basic classification scheme is drawn.

Note that figures and experimental results are shown for a wide subset of the data available in dataset 1 [48], and a subset of the data in dataset 2 [51]; the behaviour of the proposed DTW-based averaging scheme holds for other pieces of data in the datasets analysed, but not all the evaluations carried out are presented in this manuscript.

**Fig. 6.** Comparison between average ERP calculation methods: conventional average (dashed-green), DTW-based average (dotted-blue) and filtered DTW-based average (solid-red). Results are shown for EEG signals corresponding to channel 2 under the spoken-tools modality-category in [48].

### 4.1. Behaviour and performance of proposed DTW-based averaging scheme

Different analyses are carried out to illustrate the performance of the proposed DTW-based averaging scheme for EEG signals. For comparison purposes, average ERP obtained with: conventional average (baseline), DTW-based average, and filtered DTW-based average will be shown. These analyses are performed using dataset 1, described in Section 3.1.

Channels shown in Fig. 1, for subject 02, have been considered in the experiments shown to illustrate the behaviour of the proposed scheme. Note that the objective is not to identify electrodes or differences in the signal or their specific shape at the electrodes, but to focus on the behaviour of the proposed DTW-based averaging scheme.

A qualitative illustration of performance of the proposed scheme is shown in Fig. 6: average ERP waveforms of channel 2 for the spoken-tools samples obtained by using the three considered schemes are compared: in dashed-green line, the conventional average ERP is depicted, it shows clear amplitude attenuation with respect to the other approaches (and, also, with respect to the individual waveforms), however the range of its spectral content remains unaltered with respect to the original signals due to the linearity of the processing scheme; in dotted-blue line, the DTW-based average ERP waveform is shown, in this case, the signals display larger amplitude variations, and peaky resemblance that reflect the presence of high frequency components coming from the non-linear processing scheme, also note that averaging numerous trials diminishes the importance of spurious deviations that appear in each warped signal (see Fig. 2); finally, the outcome of the filtered DTW-based average scheme is drawn in solid-red line, this signal mostly maintains the amplitude, and shape of the waveform of the unfiltered version, but it is free from spurious frequency components.

Fig. 7 shows a different approach to illustrate the behaviour, and differences between the considered averaging schemes; recall that most other DTW based schemes do not allow simple signal averaging [31], and are not meant to build signal models [16,25,26]. In this figure, channel 14, corresponding to spoken modality and tools category [48] has been considered. Average signals obtained by conventional average, DTW-based average and filtered DTW-based average are shown, from top to bottom. In each case, the average signal across trials is drawn within a shaded area that represents a separation of one standard deviation measured at each sampling instant across all the selected trials, with $\sigma$, $\sigma^w$ and $\sigma^f$ standing for the standard deviations corresponding to conventional, DTW-based and filtered DTW-based averages, respectively. Note that data normalization [17] was not necessary in order to carry out the process described.

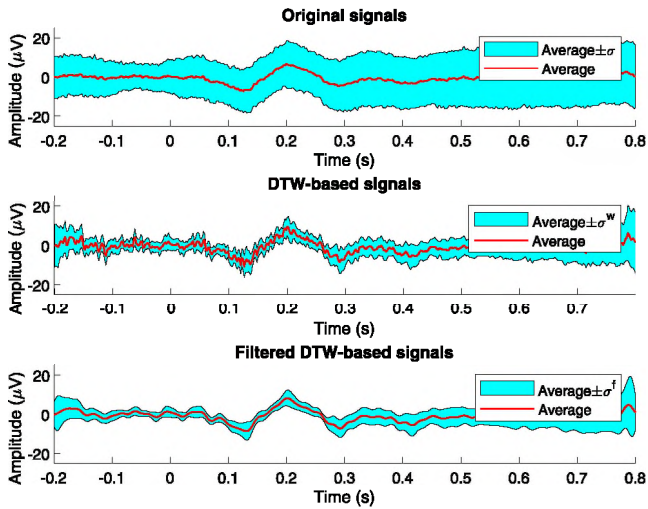

**Fig. 7.** Illustration of the average and sample-wise standard deviation of signal excerpts with respect to their average across trials for conventional, DTW-based and filtered DTW-based average, for channel 14. Data corresponds to spoken modality and tools category [48]. Average signal (in red) is drawn within a shaded area that represents a separation of one standard deviation measured (in light blue) at each sampling instant across all the trials.

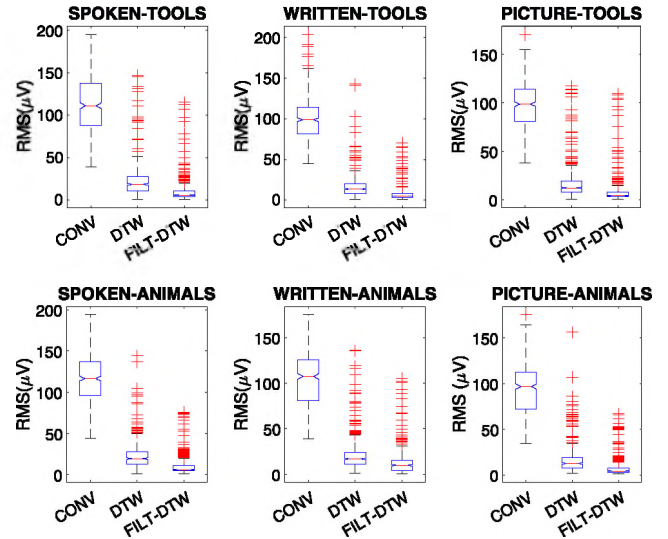

**Fig. 8.** Boxplot representation of RMS (µV) of each trial in $S_2$ with respect to conventional, DTW-based, and filtered DTW-based averages for channel 41 (see Eq. (4)). Data are shown for each modality and category of [48].

In Fig. 7, it can be observed that the standard deviation decreases significantly when DTW-based average is employed, and also EEG components seem more clearly defined; specifically, the component at 200 ms, that may be related to the processing of spoken language, is more clearly distinguished. Note that the location of channel 14 is close to the parietal area of the brain, where stimuli related to senses, in addition to attention, are perceived and processed [63]. However, the result in this case shows noisy resemblance because of undesired high frequency artefacts introduced by the DTW stage. This issue disappears when using the proposed filtered DTW-based averaging to keep the frequency content of warped signals within the range of the original ones (up to 30 Hz in our example), before performing the averaging step. Note that, also, the magnitude of standard deviation is slightly reduced with respect to the unfiltered version.

Observe that, in both DTW-based schemes, the standard deviation is larger both at the beginning and at the end of excerpts. This can be due to the constraints imposed by the DTW scheme, according to which the minimum-cost path between the aligned signal and the reference starts with the first sample in both signals, and concludes with the last sample in both signals too, and the length constraint imposed by our scheme to allow averaging. These constraints impair the performance of our schemes near the edges of signal excerpts. Recall that other DTW-based schemes need to impose other restrictions on the temporal variable [20].

Now, in order to draw proper quantitative measures, and comparison on the performance of the proposed scheme, the samples corresponding to each channel, modality and category will be split into two even subsets: $S_1$ and $S_2$. One of them, $S_1$, will be used to obtain average signals (conventional: $r[n]$, DTW-based: $r^w[n]$ and filtered DTW-based: $r^f[n]$), and the rest of trials, $s_t[n] \in S_2$, will be used to obtain measures.

RMS values are obtained for the differences between each average, $r[n]$, $r^w[n]$ and $r^f[n]$, obtained with the data in $S_1$ and the trials $s_t[n] \in S_2$ for the same channel, modality and category according to Eq. (4) and the descriptions in Section 3.5.

A sample of the boxplots found are drawn in Fig. 8. In the boxplots, the central red mark represents the median, and the upper and lower edges of each blue box draw the 75th and 25th percentiles, respectively. The outer black lines are considered the most extreme values, while red crosses represent outliers.

Observe the difference between the results obtained with conventional average and the ones found with DTW-based and filtered DTW-based averages: the median of RMS values corresponding to both DTW-based averages is notably smaller, which indicates better (lower) error in the approximation of signals by using the corresponding DTW-based averages. We consider that this is due to the diminishing of the effect of jitter, and latency jitter attained by the DTW-based process proposed. We think this allows a better construction of the representative average signal of the experiment carried out. Furthermore, the width between the 25th and 75th percentiles is also consistently smaller in the proposed DTW-based schemes.

Finally, it can be observed that RMS values of the proposed filtered DTW-based averaging scheme are slightly smaller than the ones of the DTW scheme, which should be expected due to the effect of filtering. We consider that the representation attained by this method is a better portrayal of the signals corresponding to each case of analysis. Note that the behaviour described is found not only for the data shown in Fig. 8, but for all the data analysed from [48].

Maximum absolute difference (MAD) [59,60] values, defined according to Eq. (5) in Section 3.5, are obtained between the considered average signals calculated with samples in $S_1$, and the corresponding trials in $S_2$.

MAD values corresponding to channel 22 of EEG data from [48] for each modality and category, and for the three averaging schemes considered, are represented in Fig. 9 in boxplots.

Results show a behaviour similar to the one found for the RMS measure. MAD median value in the DTW-based scheme is smaller than in the conventional one, and even smaller in the case of the filtered DTW-based scheme. These results indicate that even the individual processed samples are closer to their corresponding averages than the original samples to their conventional average. Furthermore, recall that we are considering the whole trials, including the pre-stimulus and tail, where larger difference to the average is expected, as illustrated by Fig. 7.

Regarding the analysis of ERP components to assess the performance of the proposed scheme, the criteria indicated in Section 3.5 are followed. We have considered the P200 component related to auditory stimuli (spoken), and selected the data corresponding to channel 2 in [48], taking into account [61].

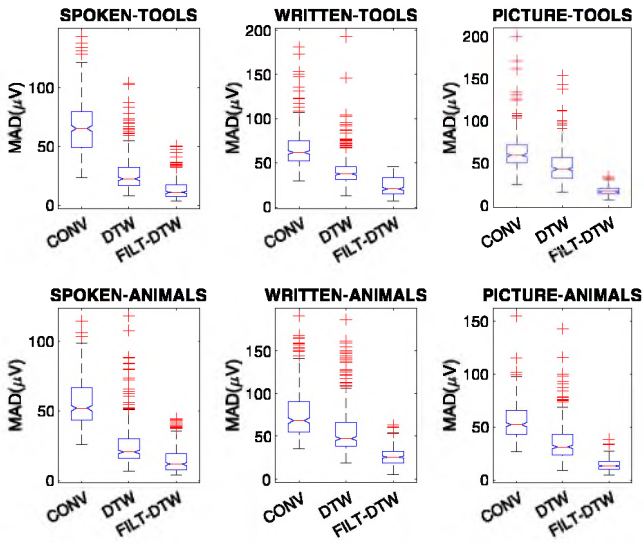

**Fig. 9.** Boxplot representation of MAD (μV) values for conventional average, DTW-based average, and filtered DTW-based average, for EEG signals of channel 22 in [48], for each modality and category.

**Table 1**
Behaviour of the parameters of P200 component in channel 2 related to spoken stimuli in [48] for the averaging schemes: conventional (Conv.), DTW-based, and filtered DTW-based (F. DTW-based).

| Delay ($s$) | Conv. | DTW-based | F. DTW-based |
|---|---|---|---|
| Mean | 0,2172 | 0,2168 | 0,2142 |
| STD | 0,0010 | 0,0025 | 0,0015 |
| V | 0,0046 | 0,0115 | 0,0070 |
| Median | 0,2180 | 0,2180 | 0,2140 |
| Q25 | 0,2160 | 0,2160 | 0,2140 |
| Q75 | 0,2180 | 0,2180 | 0,2160 |
| Max. | 0,2180 | 0,2180 | 0,2160 |
| Min. | 0,2160 | 0,2100 | 0,2120 |
| **Peak** (μV) | **Conv.** | **DTW-based** | **F. DTW-based** |
| Mean | 11,9588 | 17,1387 | 15,0694 |
| STD | 0,2351 | 0,5632 | 0,2464 |
| V | 0,0197 | 0,0329 | 0,0164 |
| Median | 11,9116 | 17,3266 | 15,0608 |
| Q25 | 11,8235 | 16,5651 | 14,8958 |
| Q75 | 12,1861 | 17,5198 | 15,2430 |
| Max. | 12,3899 | 17,8824 | 15,5116 |
| Min. | 11,6523 | 16,2106 | 14,7508 |
| **Amplitude** (μV) | **Conv.** | **DTW-based** | **F. DTW-based** |
| Mean | 20,1847 | 30,1073 | 26,2166 |
| STD | 0,3800 | 0,7176 | 0,4611 |
| V | 0,0188 | 0,0238 | 0,0176 |
| Median | 20,2532 | 30,1730 | 26,1865 |
| Q25 | 19,8003 | 29,5258 | 25,8140 |
| Q75 | 20,4004 | 30,7161 | 26,6602 |
| Max. | 20,9124 | 30,9882 | 26,8685 |
| Min. | 19,6506 | 29,0947 | 25,5681 |

However, note that now the evaluation scheme is different, we will observe the behaviour of measured P200 delay, peak and amplitude obtained by using the three schemes considered. To this end, a k-fold cross-validation scheme is employed building k=10 even folds, in this scenario, one of them is left out whereas the remaining folds are employed at each evaluation step. After delay, peak and amplitude are obtained, their mean, standard deviation, coefficient of variation, $V$, median, 25th and 75th quartiles, maximum and minimum values are calculated for each averaging scheme. Data are shown in Table 1.

In Table 1, it can be observed that mean and median values of the delay of P200 for the filtered DTW-based scheme under the evaluation framework exposed are a bit lower than in the conventional and DTW-based ones: the difference is no larger than 3.5 ms in the experiments carried out. This difference (or the reverse difference, when present) should be due to the presence of some large peaks at high (low) delay values that are filtered out before averaging in the filtered DTW-based scheme, while they remain in the other two schemes. Also note that although the coefficient of variation $V$ is larger in the DTW-based schemes, specially in the unfiltered case, its value remains very low [62].

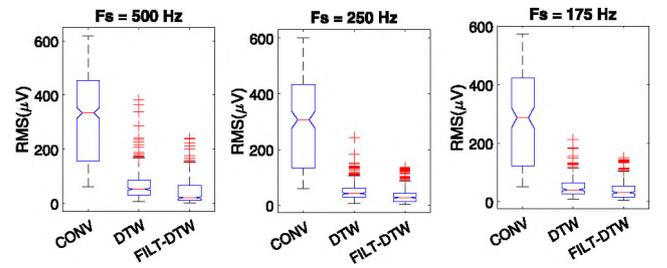

**Fig. 10.** RMS (μV) comparison for channel 33, and for the written-modality, and tools-category with various sampling rates: $F_s = 500$ Hz, $F_s = 250$ Hz and $F_s = 175$ Hz.

Regarding the peak and amplitude measures, both DTW-based schemes (unfiltered and filtered) show larger values than the conventional one. The median (and mean) value of P200 peak and amplitude for the (unfiltered) DTW-based scheme is 45.46% (43.31%) and 48.98% (49.16%) larger than with the conventional scenario, respectively. Nevertheless, roughness considerations should be taken into account that advice against its utilization, as previously described (see Fig. 7).

In the case of the filtered DTW-based scheme, median and mean values of both P200 peak and amplitude are also larger than with the conventional scenario, though smaller than in the unfiltered DTW-based case, as expected. Specifically, P200 median (and mean) peak and amplitude are 26.44% (26.01%) and 29.30% (29.88%) larger, respectively, with the filtered DTW-based scheme than with the conventional averaging method; additionally, in this case, roughness considerations should favour the choice of the filtered DTW-based scheme versus the others to define the pattern of selected components.

Also, observe that the coefficient of variation of peak and amplitude measures remains low in the DTW-based schemes. Moreover, in the case of the filtered DTW-based scheme, $V$ becomes even smaller than in the conventional scenario, which is an indicator of smaller variability in the data [62].

Finally, sensitivity to sampling rate has also been considered. To illustrate the behaviour found, RMS for channel 33 in [48] for the written-tools scenario, and three diverse sampling rates is shown in Fig. 10. Note that in this case, the whole data available of the selected modality-category are employed to obtain the results shown. Resampling was carried out by using Fieldtrip Toolbox [56]. In addition to the original sampling rate: $F_s = 500$ Hz, the following sampling rates are considered in this new boxplot: $F_s = 250$ Hz and $F_s = 175$ Hz.

It must be observed that the trend of the behaviour of RMS is the same for the three considered schemes at the different sampling rates. This behaviour holds for other channels and cases in the dataset employed for evaluation, which seems to verify that variations in sampling rate will not alter the behaviour of the proposed scheme.

### 4.2. Classification example

Dataset 2, described in Section 3.2, is used to illustrate the utilization of the exposed DTW-based averaging scheme in a classification task; the channel located 2 cm frontal of C3, is employed (Section 3.2). Conventional averaging is employed as baseline; to this end, signal models obtained by using these two schemes are obtained for the four classes available in the dataset. In Fig. 11, these signal models are drawn using a representation like the one used in Fig. 7.

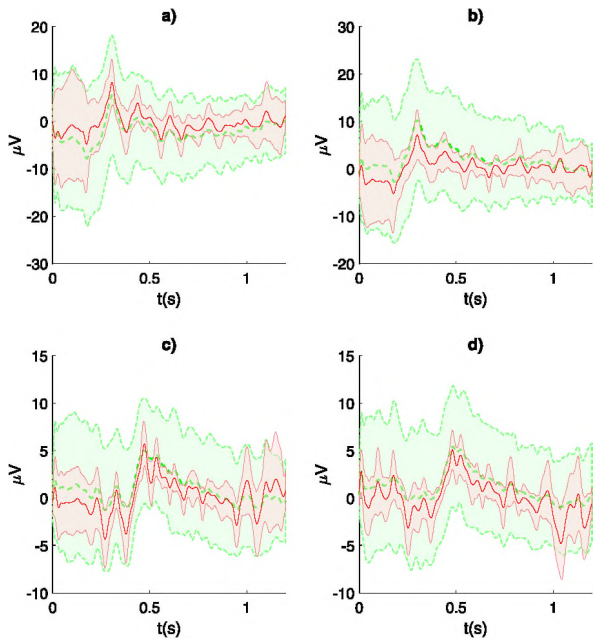

**Fig. 11.** Comparison between averages, signal models, and corresponding sample-wise standard deviation of signal excerpts obtained by using conventional average (dashed-green line, and green-shaded area surrounded by thin dashed-green line), and filtered DTW-based average (solid-red line, and red-shaded area surrounded by thin red line), for the healthy, figures: (a) and (b), and ALS subjects, figures: (c) and (d), and tasks *Up* and *Down*, left: and right figures, respectively.

**Table 2**
Average accuracy of linear SVM classification scheme using RMS of the difference between test signals and signal models of the different classes as features.

| | Averaging model | |
|---|---|---|
| | F. DTW-based | Conventional |
| Average accuracy (%) | 59.52 | 58.30 |

| True Class | Predicted Class | | | |
|---|---|---|---|---|
| | *Health Up* | *Health Down* | *ALS Up* | *ALS Down* |
| *Health Up* | 72 | 25 | 2 | 4 |
| *Health Down* | 30 | 55 | 2 | 1 |
| *ALS Up* | 3 | 4 | 31 | 29 |
| *ALS Down* | 5 | 2 | 24 | 38 |

**Fig. 12.** Sample confusion matrix: multi-class linear SVM classifier using RMS of the difference between signals, and filtered DTW-based average signal models as features.

Observe that signal models obtained by using the filtered DTW-based averaging scheme exhibit sharper peaks and troughs than signals obtained by conventional averaging, which means that the flattening effect of averaging has been indeed diminished, and also the sample-by-sample dispersion of the processed signals with respect to their average is significantly smaller. All this happens in spite of the fact that the number of trials available is significantly smaller than previously. Nevertheless, the salience of peaks and valleys in the filtered DTW-based signals, with respect to the conventional average, is not as high as previously (see Fig. 6; the reason for this can be found in the significantly smaller number of trials available in the current scenario: a large number of trials would benefit the performance of the procedure proposed).

Now, among the diverse scenarios and choices that can be considered, a simple classification task and scheme are selected: classification between *Up* and *Down* tasks for the healthy and ALS cases, on the basis of RMS of the difference between the EEG signals and the corresponding signal models. This choice is consistent with the findings shown earlier in Section 4. SVM classifier with linear kernel, using one-versus-one multi-class classification strategy, and 5-fold cross-validation is used to perform the classification of every trial into one of the four classes: Health-Up, Health-Down, ALS-up and ALS-down. The RMS of the difference between input signals, and signal models of each class obtained by filtered DTW-based/conventional averaging, see Eq. (4), is used to build a simple 4-dimensional feature vector for classification. With all this, the accuracy attained in this task is drawn in Table 2.

Note that in spite of the obvious similarities between the signals, specially in the ALS patient case, as shown in Fig. 11, the simplicity of the characterization scheme used, and the limited number of samples (after artefact removal stage, split of data into four classes for classification, and strict separation between training and test data in our cross-validation scheme, only 67 trials were available in the less populated class, see Section 3.3), the proposed scheme allows slight consistent improvement of classification accuracy in the evaluations performed. Fig. 12 illustrates this behaviour by means of a confusion matrix; this allows to observe the difficulties in the classification between *Up* and *Down* classes in both the Health and, specially, ALS cases; this is the same trend observed when conventional average is employed (the distribution of the confusion matrix is similar, with worse performance regarding the distinction between ALS cases), though scores are lower, which, taking into account the classification scheme employed, implies that signal models corresponding to different classes obtained by using the proposed scheme are more different from each other than those obtained by using conventional averaging.

It must be taken into account that, in order to perform a fair comparison, training and test samples needed to be fully separated; this means that test samples were neither part of the data subset used to build the signal models per class nor part of the data subset used, afterward, to train the classification scheme.

## 5. Conclusion

In this paper, a new scheme for trial averaging in the context of ERP signal analysis is proposed. This scheme is also based on averaging, though it relies on the utilization of a modified version of the well-known DTW algorithm, and the application of a filtering step to adequate warped signals to the frequency range where they should be.

The evidence provided on the behaviour of the proposed scheme suggests that it should be considered as a possible alternative to conventional averaging for lessening the influence of jitter and latency jitter, and, consequently, for the extraction of parameters for BCI or the evaluation of neurological conditions.

Nevertheless, further analyses carried out were aimed at evaluating the performance of the proposed scheme, not only regarding the sampling frequency, but also about the importance of the number of trials, influence of the non-linearity of the processing scheme, performance at the edges of signal excerpts or the gain in peak and amplitude values of ERP waves.

Finally, recall that experiments carried out were aimed at evaluating the performance of the proposed filtered DTW-based averaging scheme; the specific analysis of brain behaviour is out of the scope of this work, unlike the signal processing perspective.

**CRediT authorship contribution statement**

**Mario Molina:** Conceptualization, Methodology, Software, Investigation, writing. **Lorenzo J. Tardón:** Conceptualization, Methodology, Software, Investigation, Writing, Project administration. **Ana M. Barbancho:** Conceptualization, Methodology, Software, Investigation, Validation, Writing. **Irene De-Torres:** Conceptualization, Methodology, Investigation, Validation, Writing. **Isabel Barbancho:** Conceptualization, Methodology, Software, Investigation, Validation, Writing, Project administration.

**Declaration of competing interest**

The authors declare that they have no known competing financial interests or personal relationships that could have appeared to influence the work reported in this paper.

**Data availability**

Data are available from other researchers, as indicated in the manuscript.

**Acknowledgements**

This publication is part of project PID2021-123207NB-I00, funded by MCIN/AEI /10.13039/501100011033 / FEDER, UE. This work was partially funded by Junta de Andalucía, Proyectos de I+D+i, in the framework of Plan Andaluz de Investigación, Desarrollo e Innovación (PAIDI 2020), under Project No. PY20_00237. Funding for open access charge: Universidad de Málaga/CBUA. This work was done at Universidad de Málaga, Campus de Excelencia Internacional Andalucía Tech.

**References**

[1] Z. Leonowicz, J. Karvanen, S.L. Shishkin, Trimmed estimators for robust averaging of event-related potentials, J. Neurosci. Methods 142 (1) (2005) 17–26.

[2] D. Heo, M. Kim, J. Kim, Y.J. Choi, S.-P. Kim, The uses of brain-computer interface in different postures to application in real life, in: 2022 10th International Winter Conference on Brain-Computer Interface, BCI, 2022, pp. 1–5, http://dx.doi.org/10.1109/BCI53720.2022.9734957.

[3] Y. Wang, M. Nakanishi, D. Zhang, EEG-based brain-computer interfaces, in: X. Zheng (Ed.), Neural Interface: Frontiers and Applications, Springer Singapore, Singapore, 2019, pp. 41–65, http://dx.doi.org/10.1007/978-981-13-2050-7_2.

[4] L.M. McCane, S.M. Heckman, D.J. McFarland, G. Townsend, J.N. Mak, E.W. Sellers, D. Zeitlin, L.M. Tenteromano, J.R. Wolpaw, T.M. Vaughan, P300-based brain-computer interface (BCI) event-related potentials (ERPs): People with amyotrophic lateral sclerosis (ALS) vs. age-matched controls, Clin. Neurophysiol. 126 (11) (2015) 2124–2131, http://dx.doi.org/10.1016/j.clinph.2015.01.013, URL: https://www.sciencedirect.com/science/article/pii/S1388245715000067X.

[5] M.D. Burns, N. Bigdely-Shamlo, N.J. Smith, K. Kreutz-Delgado, S. Makeig, Comparison of averaging and regression techniques for estimating event related potentials, in: 2013 35th Annual International Conference of the IEEE Engineering in Medicine and Biology Society, EMBC, IEEE, 2013, pp. 1680–1683.

[6] S.J. Luck, An Introduction to the Event-Related Potential Technique, MIT Press, Cambridge, MA, 2005.

[7] S.J. Luck, An Introduction to the Event-Related Potential Technique, MIT Press, 2014.

[8] L. Hu, Z. Zhang, EEG Signal Processing and Feature Extraction, Springer, 2019, http://dx.doi.org/10.1007/978-981-13-9113-2.

[9] A. de Cheveigné, D.D. Wong, G.M. Di Liberto, J. Hjortkjaer, M. Slaney, E. Lalor, Decoding the auditory brain with canonical component analysis, NeuroImage 172 (2018) 206–216.

[10] G.A. Light, L.E. Williams, F. Minow, J. Sprock, A. Rissling, R. Sharp, N.R. Swerdlow, D.L. Braff, Electroencephalography (EEG) and event-related potentials (ERPs) with human participants, Curr. Protocols Neurosci. 52 (1) (2010) 6.25.1–6.25.24, http://dx.doi.org/10.1002/0471142301.ns0625s52, URL: https://currentprotocols.onlinelibrary.wiley.com/doi/abs/10.1002/0471142301.ns0625s52.

[11] O. Tandon, A.S. Mahajan, Averaged evoked potentials: Event related potentials (ERPs) and their applications, Indian J. Physiol. Pharmacol. 43 (1999) 425–434.

[12] V.J. Samar, K.P. Swartz, M.R. Raghuveer, Multiresolution analysis of event-related potentials by wavelet decomposition, Brain Cognit. 27 (3) (1995) 398–438.

[13] J. Li, Z.L. Yu, Z. Gu, M. Tan, Y. Wang, Y. Li, Spatial–temporal discriminative restricted Boltzmann machine for event-related potential detection and analysis, IEEE Trans. Neural Syst. Rehabil. Eng. 27 (2) (2019) 139–151.

[14] A. Mouraux, G.D. Iannetti, Across-trial averaging of event-related EEG responses and beyond, Magn. Reson. Imaging 26 (7) (2008) 1041–1054.

[15] A.H. Zisk, S.B. Borgheai, J. McLinden, S.M. Hosni, R.J. Deligani, Y. Shahriari, P300 latency jitter and its correlates in people with amyotrophic lateral sclerosis, Clin. Neurophysiol. 132 (2) (2021) 632–642.

[16] M. Ahmadi, M.A. Schoenfeld, S.A. Hillyard, R.Q. Quiroga, A simple metric to study the mechanisms generating event-related potentials, J. Neurosci. Methods 360 (2021) 109230.

[17] K. Wang, H. Begleiter, B. Porjesz, Warp-averaging event-related potentials, Clin. Neurophysiol. 112 (10) (2001) 1917–1924.

[18] G. Ouyang, W. Sommer, C. Zhou, Reconstructing ERP amplitude effects after compensating for trial-to-trial latency jitter: A solution based on a novel application of residue iteration decomposition, Int. J. Psychophysiol. 109 (2016) 9–20, http://dx.doi.org/10.1016/j.ijpsycho.2016.09.015, URL: https://www.sciencedirect.com/science/article/pii/S0167876016307103.

[19] C.D. Woody, Characterization of an adaptive filter for the analysis of variable latency neuroelectric signals, Med. Biol. Eng. 5 (1967) 539–554.

[20] H. Gibbons, J. Stahl, Response-time corrected averaging of event-related potentials, Clin. Neurophysiol. 118 (1) (2007) 197–208.

[21] S. Casarotto, A. Bianchi, S. Cerutti, G. Chiarenza, Dynamic time warping in the analysis of event-related potentials, IEEE Eng. Med. Biolo. Mag. 24 (1) (2005) 68–77, http://dx.doi.org/10.1109/MEMB.2005.1384103.

[22] M. Kotas, T. Pander, J.M. Leski, Averaging of nonlinearly aligned signal cycles for noise suppression, Biomed. Signal Process. Control 21 (2015) 157–168, http://dx.doi.org/10.1016/j.bspc.2015.06.003, URL: https://www.sciencedirect.com/science/article/pii/S1746809415001056.

[23] T. Fukami, J. Watanabe, F. Ishikawa, Robust estimation of event-related potentials via particle filter, Comput. Methods Programs Biomed. 125 (2016) 26–36.

[24] H.-C. Huang, B. Jansen, EEG waveform analysis by means of Dynamic Time-Warping, Int. J. Bio-Med. Comput. 17 (2) (1985) 135–144.

[25] Q. Li, G.D. Clifford, Dynamic Time Warping and Machine Learning for signal quality assessment of pulsatile signals, Physiol. Measur. 33 (9) (2012) 1491.

[26] T. Yamauchi, K. Xiao, C. Bowman, A. Mueen, Dynamic time warping: A single dry electrode EEG study in a self-paced learning task, in: International Conference on Affective Computing and Intelligent Interaction, ACII, 2015, pp. 56–62, http://dx.doi.org/10.1109/ACII.2015.7344551.

[27] Y.-T. Liu, Y.-A. Zhang, M. Zeng, Adaptive global time sequence averaging method using dynamic time warping, IEEE Trans. Signal. Process. 67 (8) (2019) 2129–2142, http://dx.doi.org/10.1109/TSP.2019.2897958.

[28] F. Petitjean, A. Ketterlin, P. Gançarski, A global averaging method for dynamic time warping, with applications to clustering, Pattern Recognit. 44 (3) (2011) 678–693.

[29] J.W. Li, S. Barma, S.H. Pun, F. Chen, C. Li, M.T. Li, P.K. Wang, M.I. Vai, P.U. Mak, EEG-based emotion recognition using similarity measure of brain rhythm sequencing, in: 43rd Annual International Conference of the IEEE Engineering in Medicine & Biology Society, EMBC, 2021, pp. 1–4, http://dx.doi.org/10.1109/EMBC46164.2021.9629520.

[30] J.W. Li, S. Barma, P.U. Mak, F. Chen, C. Li, M.T. Li, M.I. Vai, S.H. Pun, Single-channel selection for EEG-based emotion recognition using brain rhythm sequencing, IEEE J. Biomed. Health Inf. 26 (6) (2022) 2493–2503.

[31] W. Zhao, Z. Xu, W. Li, W. Wu, Modeling and analyzing neural signals with phase variability using Fisher-Rao registration, J. Neurosci. Methods 346 (2020) 108954.

[32] S. Sur, V.K. Sinha, Event-related potential: An overview, Ind. Psychiatry J. 18 (2009) 70–73.

[33] A.B. Usakli, Improvement of EEG signal acquisition: An electrical aspect for state of the art of front end, Comput. Intell. Neurosci. 2010 (2010).

[34] D.E.J. Linden, The P300: Where in the brain is it produced and what does it tell us? Neuroscientist 11 (6) (2005) 563–576, http://dx.doi.org/10.1177/1073858405280524.

[35] R. Verleger, P. Jaśkowski, E. Wascher, Evidence for an integrative role of P3b in linking reaction to perception, J. Psychophysiol. 19 (3) (2005) 165–181, http://dx.doi.org/10.1027/0269-8803.19.3.165.

[36] X. Yu, The Impact of Latency Jitter on the Interpretation of P300 in the Assessment of Cognitive Function (Ph.D. thesis), University of South Florida, 2016, URL: https://www.semanticscholar.org/paper/The-Impact-of-Latency-Jitter-on-the-Interpretation-Yu/511a1ac2b73c426236ce737cd4eb36b68e8ec70e, visited on 2022-03-03.

[37] K.B. Walhovd, H. Rosquist, A.M. Fjell, P300 amplitude age reductions are not caused by latency jitter, Psychophysiology 45 (4) (2008) 545–553.

[38] R. Freunberger, W. Klimesch, M. Doppelmayr, Y. Höller, Visual P2 component is related to theta phase-locking, Neurosci. Lett. 426 (3) (2007) 181–186, http://dx.doi.org/10.1016/j.neulet.2007.08.062, URL: https://www.sciencedirect.com/science/article/pii/S0304394007009482.

[39] X. Du, F.-S. Choa, A. Summerfelt, L. Rowland, J. Chiappelli, P. Kochunov, N100 as a generic cortical electrophysiological marker based on decomposition of TMS-evoked potentials across five anatomic locations, Exp. Brain Res. 235 (2017) http://dx.doi.org/10.1007/s00221-016-4773-7.

[40] W. Giroldini, L. Pederzoli, M. Bilucaglia, S. Melloni, P. Tressoldi, A new method to detect event-related potentials based on Pearson's correlation, EURASIP J. Bioinform. Syst. Biol. 2016 (2016) http://dx.doi.org/10.1186/s13637-016-0043-z.

[41] J.G. Murray, G. Ouyang, D.I. Donaldson, Compensation of trial-to-trial latency jitter reveals the parietal retrieval success effect to be both variable and thresholded in older adults, Front. Ag. Neurosci. 11 (2019) 179, http://dx.doi.org/10.3389/fnagi.2019.00179, URL: https://www.frontiersin.org/article/10.3389/fnagi.2019.00179.

[42] D.J. Berndt, J. Clifford, Using Dynamic Time Warping to find patterns in time series, in: KDD Workshop, Vol. 10, no. 16, Seattle, WA, USA:, 1994, pp. 359–370.

[43] D. Ellis, Dynamic Time Warping (DTW) in MATLAB, 2022, URL: https://www.ee.columbia.edu/~dpwe/resources/matlab/dtw/. (Accessed on 03 March 2022).

[44] M. Müller, Information Retrieval for Music and Motion, Springer Berlin Heidelberg, Berlin, Heidelberg, 2007, pp. 69–84.

[45] B.H. Ricardo Portilla, D. Lee, Understanding dynamic time warping, 2019, URL: https://databricks.com/blog/2019/04/30/understanding-dynamic-time-warping.html, visited on 2022-03-03.

[46] P. Senin, Dynamic Time Warping Algorithm Review, Vol. 855, no. 1–23, Information and Computer Science Department University of Hawaii at Manoa Honolulu, USA, 2008, p. 40.

[47] B.J. Jain, D. Schultz, Optimal warping paths are unique for almost every pair of time series, 2017, CoRR abs/1705.05681, URL: http://arxiv.org/abs/1705.05681, arXiv:1705.05681.

[48] I. Simanova, M. van Gerven, R. Oostenveld, P. Hagoort, Identifying object categories from event-related EEG: Toward decoding of conceptual representations, PLoS One 5 (12) (2010) 1–12, http://dx.doi.org/10.1371/journal.pone.0014465.

[49] BIDS Community, Brain imaging data structure, 2022, URL: https://bids.neuroimaging.io/, visited on 2022-03-03.

[50] S.J. Luck, A.X. Stewart, A.M. Simmons, M. Rhemtulla, Standardized measurement error: A universal metric of data quality for averaged event-related potentials, Psychophysiology 58 (6) (2021) e13793.

[51] N. Birbaumer, H. Flor, N. Ghanayim, T. Hinterberger, I. Iverson, E. Taub, B. Kotchoubey, A. Kbler, J. Perelmouter, EEG dataset of slow cortical potentials, 2023, URL: https://www.kaggle.com/datasets/towsifahamed/eeg-dataset-of-slow-cortical-potentials. (Accessed on 05 May 2023).

[52] N. Birbaumer, H. Flor, N. Ghanayim, T. Hinterberger, I. Iverson, E. Taub, B. Kotchoubey, A. Kbler, J. Perelmouter, A brain-controlled spelling device for the completely paralyzed, Nature398 (2001) 297–298.

[53] M.C. Kiernan, S. Vucic, B.C. Cheah, M.R. Turner, A. Eisen, O. Hardiman, J.R. Burrell, M.C. Zoing, Amyotrophic lateral sclerosis, The Lancet 377 (9769) (2011) 942–955.

[54] PychLab, PsyLab EEG8 amplifier, 2023, URL: http://www.psychlab.com/EEG_8_amplifier.html. (Accessed on 30 May 2023).

[55] G.H. Klem, The ten-twenty electrode system of the international federation. The international federation of clinical neurophysiology, Electroencephalogr. Clin. Neurophysiol. Suppl. 52 (1999) 3–6.

[56] R. Oostenveld, P. Fries, E. Maris, J.-M. Schoffelen, FieldTrip: Open source software for advanced analysis of MEG, EEG, and invasive electrophysiological data, Comput. Intell. Neurosci. 2011 (2011).

[57] Y.-P. Lin, P. Vaidyanathan, A Kaiser window approach for the design of prototype filters of cosine modulated filterbanks, IEEE Signal Process/ Lett. 5 (6) (1998) 132–134.

[58] J.S. Garofolo, L.F. Lamel, W.M. Fisher, J.G. Fiscus, D.S. Pallett, N.L. Dahlgren, DARPA TIMIT Acoustic Phonetic Continuous Speech Corpus CDROM, NIST, 1993.

[59] J.M. Leski, Robust weighted averaging [of biomedical signals], IEEE Trans. Biomed. Eng. 49 (8) (2002) 796–804.

[60] K. Kotowski, K. Stapor, J. Leski, Improved robust weighted averaging for event-related potentials in EEG, Biocybern. Biomed. Eng. 39 (4) (2019) 1036–1046.

[61] C.-L. Shen, T.-L. Chou, W.-S. Lai, M.H. Hsieh, C.-C. Liu, C.-M. Liu, H.-G. Hwu, P50, N100, and P200 auditory sensory gating deficits in Schizophrenia patients, Front. Psychiatry 11 (2020) http://dx.doi.org/10.3389/fpsyt.2020.00868, URL: https://www.frontiersin.org/articles/10.3389/fpsyt.2020.00868.

[62] A.G. Bedeian, K.W. Mossholder, On the use of the coefficient of variation as a measure of diversity, Organ. Res. Methods 3 (3) (2000) 285–297.

[63] A. Estévez-González, C. García-Sánchez, C. Junqué, La atención: una compleja función cerebral, Revista de neurología 25 (148) (1997) 1989–1997.